# Decrease of upper critical field with underdoping in cuprate superconductors

J. Chang[1,†], N. Doiron-Leyraud[1], O. Cyr-Choinière[1], F. Laliberté[1], E. Hassinger[1], J.-Ph. Reid[1], R. Daou[1,‡], S. Pyon[2], T. Takayama[2], H. Takagi[2,3] & Louis Taillefer[1,4]

*1 Département de physique and RQMP, Université de Sherbrooke, Sherbrooke, Québec J1K 2R1, Canada*

*2 Department of Advanced Materials, University of Tokyo, Kashiwa 277-8561, Japan*

*3 RIKEN (The Institute of Physical and Chemical Research), Wako, 351-0198, Japan*

*4 Canadian Institute for Advanced Research, Toronto, Ontario M5G 1Z8, Canada*

**The transition temperature $T_c$ of cuprate superconductors falls when the doping $p$ is reduced below a certain optimal value. It is unclear whether this fall is due to strong phase fluctuations[1] or to a decrease in the pairing gap. Different interpretations of photoemission data disagree on the evolution of the pairing gap[2,3,4,5] and different estimates of the upper critical field $H_{c2}$ are in sharp contradiction[6,7]. Here we resolve this contradiction by showing that superconducting fluctuations in the underdoped cuprate $La_{1.8-x}Eu_{0.2}Sr_xCuO_4$, measured via the Nernst effect, have a characteristic field scale that falls with underdoping. The critical field $H_{c2}$ dips at $p = 0.11$, showing that superconductivity is weak where stripe order is strong[8]. In the archetypal cuprate superconductor**

---

[†] Present address : Institut de la matière complexe, École Polytechnique Fédérale de Lausanne (EPFL), CH-1015 Lausanne, Switzerland & Paul Scherrer Institut, CH-5232 Villigen, Switzerland.

[‡] Present address : Laboratoire de Cristallographie et Sciences des Matériaux, CNRS (UMR6508), Caen 14050, France.

**YBa$_2$Cu$_3$O$_y$, $H_{c2}$ extracted from other measurements[6,9,10,11] has the same doping dependence, also with a minimum at $p = 0.11$, again where stripe order is present[11,12]. We conclude that competing states such as stripe order weaken superconductivity and this, rather than phase fluctuations, causes $T_c$ to fall as cuprates become underdoped.**

Two paradigms have been proposed to account for the dome-like region of superconductivity in the temperature-doping phase diagram of cuprate superconductors[13]. In the first, the amplitude of the superconducting order parameter grows monotonically as the doping $p$ is reduced, but its phase is increasingly disordered[1], causing $T_c$ to fall at low $p$. The signature of this scenario is strong phase fluctuations and a gap above $T_c$ in the underdoped regime. In the second paradigm, the fall of $T_c$ at low $p$ is due to the onset of a state that competes with superconductivity. The signature of this scenario is a small superconducting gap and a small $H_{c2}$ in the underdoped regime.

Whether strong phase fluctuations or a decrease in the pairing gap is causing $T_c$ to fall in underdoped cuprates is currently an open question. Different interpretations of photoemission data disagree on the evolution of the pairing gap[2,3,4,5] and different estimates of the upper critical field $H_{c2}$ are in sharp contradiction[6,7]. The Nernst signal observed above $T_c$ in underdoped cuprates has been attributed to superconducting fluctuations[7,14,15], and because it persists up to temperatures several times $T_c$, it was deemed incompatible with the standard Gaussian fluctuations of the superconducting order parameter. It was attributed instead to vortex-like excitations in a phase-fluctuating superconductor[14,15] with a non-zero pairing amplitude above $T_c$. The critical field $H_{c2}$ deduced from the Nernst data on cuprates such as Bi$_2$Sr$_2$CaCu$_2$O$_{8+\delta}$ (Bi-2212) was reported to increase with underdoping[7], even though $T_c$ falls. As shown in Fig. 1, this is in striking contrast with the rapid drop in $H_{c2}$ deduced from a Gaussian analysis of fluctuations in the magneto-conductivity of YBa$_2$Cu$_3$O$_y$ (YBCO) (6).

Here we re-examine the Nernst effect in cuprates with a study of La$_{1.8-x}$Eu$_{0.2}$Sr$_x$CuO$_4$ (Eu-LSCO), an underdoped cuprate in which the ratio of superconducting ($N_{sc}$) to quasiparticle ($N_{qp}$) contributions to the Nernst signal $N$ is exceptionally large – at least 100 times larger than in previous studies of superconducting fluctuations in cuprates (see Table S2). Because of its low $T_c$, we could determine the quasiparticle background $N_{qp}(T)$ in Eu-LSCO by fully suppressing superconductivity with a magnetic field (Fig. S1). The large signal-to-background ratio allows us to reliably track $N_{sc}$ up to high temperature, namely up to $\varepsilon \equiv (T - T_c) / T_c \approx 5$, compared to a typical upper limit of $\varepsilon \approx 0.5$. As we shall see, this gives us access to a regime where the complicating effects of paraconductivity are negligible.

In Fig. 2a, $N_{sc}$ is plotted vs magnetic field $H$ for different temperatures above $T_c$, for Eu-LSCO at a doping $p = 0.11$. $N_{sc}$ increases linearly at low $H$, peaks at a field $H^*$ and then decreases monotonically at high $H$, just as in the conventional superconductor Nb$_{0.15}$Si$_{0.85}$ (refs. 16, 17) (Fig. S4). The peak field $H^*$, called the "ghost critical field" (ref. 17), is plotted vs $\varepsilon$ in Fig. 2b. It obeys $H^* = H_{c2}^* \ln(T / T_c)$ from $\varepsilon \approx 0.5$ to $\varepsilon \approx 5$, where $H_{c2}^*$ is a field scale whose relation to the $T = 0$ upper critical field $H_{c2}$ is discussed below.

Below $\varepsilon \approx 0.5$, $H^*$ deviates from $\ln(T / T_c)$, and remains finite as $\varepsilon \to 0$. This is because $N_{sc}(\varepsilon) = \alpha_{xy}^{sc}(\varepsilon) / \sigma(\varepsilon)$ is the ratio of two quantities[18,19,20,21] – the off-diagonal Peltier coefficient from superconducting fluctuations $\alpha_{xy}^{sc}$ and the electrical conductivity $\sigma$ – which both diverge as $\varepsilon \to 0$ (ref. 18). This causes $N_{sc}$ to saturate at low $\varepsilon$ (Fig. S5). The deviation of $H^*$ from $\ln(T / T_c)$ coincides with the onset of paraconductivity below $T \approx 6$ K $\approx 1.5\, T_c$ (see Fig. S6). Above $\varepsilon \approx 0.5$, paraconductivity is negligible (Fig. S7) and $\sigma$ reaches its (field-independent) normal-state value, at which point $N_{sc}(H) \sim \alpha_{xy}^{sc}(H)$. We make use of $H^*$ in the latter regime only.

$H^*$ also obeys $H^* = H_{c2}^* \ln(T/T_c)$ in our other Eu-LSCO samples (Table S1 and Fig. S3), with $p = 0.08$, 0.10 and 0.125 (Fig. 2c). The value of $H_{c2}^*$ extracted from the fit at each doping is plotted in Fig. 3a. Our first and main finding is this: the field scale for superconductivity, $H_{c2}^*$, decreases with underdoping, in a non-monotonic way, with a local minimum at $p = 0.11$. This result comes directly from the Nernst data, free of any model or theory. In fact, the evolution of $H_{c2}^*$ may be read off the raw $N$ vs $H$ isotherms: it is simply proportional to the field $H^*$ at which $N$ peaks for a given reduced temperature, say $T = 1.5\, T_c$ (Fig. S8).

A similar approach was used by Wang *et al.* in ref. 7: they extracted a field scale from their raw Nernst data on the cuprates $Bi_2Sr_{2-y}La_yCu_2O_6$ (Bi-2201) and Bi-2212, and found it to increase with underdoping (Figs. 1 and S10). However, because they used Nernst data at $T = T_c$, their analysis was contaminated by paraconductivity. Analysis of their data away from $T_c$ yields a field scale that decreases with underdoping, in agreement with diamagnetism data on Bi-2201 (Figs. S9 and S10). The $H$ dependence of $M_d$, the diamagnetic component of magnetization, is very similar to that of $N_{sc}$. Data on an underdoped sample of Bi-2201 with $T_c = 12$ K (ref. 22) yield a peak value $H_d^*$ that obeys $H_{c2}^* \ln(T / T_c)$ all the way from $T \approx T_c$ to $T \approx 4\, T_c$ (Fig. S9), with $H_{c2}^* \approx 19$ T. Applying the same fit to published Nernst data on a Bi-2201 sample of the same doping[7] yields the same value of $H_{c2}^*$ (Fig. S9).

Nernst data on $Nb_{0.15}Si_{0.85}$ (refs. 16,17) yield a peak field in agreement with $H^* = H_{c2}^* \ln(T/T_c)$ up to at least $5\, T_c$ (see Fig. 2c). Pourret *et al.* point out that $H^*$ separates a low-$H$ regime controlled by the temperature-dependent coherence length $\xi(T) = \xi_0 / (\ln(T/T_c))^{1/2}$ and a high-$H$ regime controlled by the magnetic length $l_B = (\hbar / 2eH)^{1/2}$ (ref. 23). They argue that $H^*$ is the field where $\xi(T) = l_B(H^*)$, so that $H^* = (\Phi_0 / 2\pi\xi_0^2) \ln(T / T_c)$, where $\Phi_0 = h / 2e$. This makes the field scale $H_{c2}^*$ equal to the $T = 0$ upper critical field $H_{c2} \equiv \Phi_0 / 2\pi\xi_0^2$. In Fig. S11, we show that the field needed to suppress superconductivity in Eu-LSCO at $T << T_c$ is roughly equal to $H_{c2}*$. This confirms experimentally that in Eu-LSCO, $H_{c2}* \approx H_{c2}$.

In YBCO, a cuprate with $T_c \approx 60$ K at $p = 0.11$, the effect of superconducting fluctuations on the in-plane electrical conductivity σ was analyzed up to $\varepsilon \approx 1$ for a range of dopings[6], using the Aslamazov-Larkin theory of Gaussian fluctuations. The only fit parameter in the theory is $\xi_0$, plotted in Fig. 3a as $H_{c2} = \Phi_0 / 2\pi\xi_0^2$ vs $p$. $H_{c2}$ is seen to have a minimum at $p = 0.11$, just as in Eu-LSCO. In Fig. 3b, we show that this value of $H_{c2}$ (obtained from fluctuations above $T_c$) is in good agreement with the value of $H_{c2}$ measured at $T << T_c$ directly by high-field transport[10,11] or estimated from the vortex core radius[9]. This is compelling evidence for the validity of Gaussian theory and for the low value of $H_{c2}$ in underdoped YBCO, with $H_{c2} \approx 30$ T at $p = 0.11$. We conclude that the upper critical field $H_{c2}$ of cuprates decreases with underdoping, in the same non-monotonic fashion in two very different materials.

We attribute this non-monotonic weakening of superconductivity to the competing effect of stripe order. Stripe order is present in Eu-LSCO above $p = 0.08$ (ref. 8). In YBCO, stripe order was recently inferred from Seebeck measurements of Fermi-surface reconstruction[11] and confirmed by high-field NMR measurements[12]. This scenario of phase competition is akin to that found in iron-based, heavy-fermion and organic superconductors, where the competing phase is spin-density-wave order. In YBCO at lower doping ($p < 0.08$), the rapid drop in $T_c$ and $H_{c2}$ (Fig. 3b) may be due to other phases, such as spin-density-wave order below $p \approx 0.08$ (ref. 24) and antiferromagnetism below $p \approx 0.05$. At low doping, the approach to the Mott insulator may also play a role.

We now compare our data with the theory of Gaussian fluctuations[18,19,20]. The calculated curve of $\alpha_{xy}^{sc}$ vs $H$ (ref. 20) is in excellent agreement with the measured curve of $N_{sc}$ vs $H$ (Fig. S12). The peak field in $\alpha_{xy}^{sc}$ vs $H$ increases with temperature roughly as $H^* \sim \ln(T/T_c)$ (Fig. S13), with a prefactor that is proportional to $1/\xi_0^2$ (Fig. S13). In the $H = 0$ limit, theory predicts[19,20]: $\nu_{sc}\,\sigma = \alpha_{xy}^{sc}/H \sim \xi_0^2/T\ln(T/T_c)$. In Fig. 4a, the Nernst coefficient $\nu\ (\equiv N/H)$ of Eu-LSCO at $p = 0.11$ is plotted vs $H$ for different temperatures above $T_c$. Its value in the $H = 0$ limit, $\nu_0 \equiv \nu(H \rightarrow 0)$, is plotted in Fig. 4b as a function of $\varepsilon$. In Fig. 4c, the data are seen to follow the theoretical temperature dependence precisely, from 1.02 $T_c$ up to at least 5 $T_c$, as previously found[19,20] in $Nb_{0.15}Si_{0.85}$ (refs. 16,17). We conclude that our Nernst data on Eu-LSCO are consistent with several non-trivial features of Gaussian theory. This validates the earlier use of Gaussian theory to analyze conductivity data[6] (in a context of much smaller signal-to-background ratio; see Table S2). Note that the quantum oscillations observed in YBCO at $p \sim 0.1$ (ref. 25) are consistent with Fermi-liquid theory[26], the framework on which Gaussian theory is based. Agreement with Gaussian theory and consistency of the different measures of $H_{c2}$ indicate that the superconducting fluctuations in these cuprates are controlled entirely by the coherence length, and there is only one temperature scale, $T_c$, and one field scale, $H_{c2}$, for superconductivity.

## Acknowledgements

We thank H. Aubin, K. Behnia, A. M. Finkel'stein, V. Galitski, S. A. Kivelson, K. Michaeli, A. J. Millis, M. R. Norman, M. Serbyn, M. A. Skvortsov, A.-M. Tremblay, D. van der Marel, A. Varlamov, and S. Weyerneth for fruitful discussions. We thank S.Y. Li for the resistivity data on Nd-LSCO (Fig. S11) and J. Corbin for his assistance with the experiments. We thank K. Michaeli for her unpublished calculations in Figs. S12 and S13. We thank the LNCMI for access to a high-field magnet allowing us to get data at 28 T (Fig. S1). J.C. was supported by a Fellowship from the FQRNT and the Swiss SNF. E.H. was supported by a Fellowship from the FQRNT and a Junior Fellowship from the Canadian Institute for Advanced Research (CIFAR). L.T. acknowledges support from CIFAR and funding from NSERC, FQRNT, the Canada Foundation for Innovation, and a Canada Research Chair.

## Author contributions

J.C. initiated the project; J.C., N.D.-L., O.C.-C., F.L., E.H., J.-Ph.R. and R.D. performed the Nernst measurements in Sherbrooke; J.C., N.D.-L. and F.L. performed the Nernst measurements at the LNCMI in Grenoble; S.P., T.T. and H.T. prepared the Eu-LSCO samples and measured their resistivity; J.C., N.D.-L. and L.T. analyzed the data; J.C., N.D.-L. and L.T. wrote the manuscript; L.T. supervised the project.

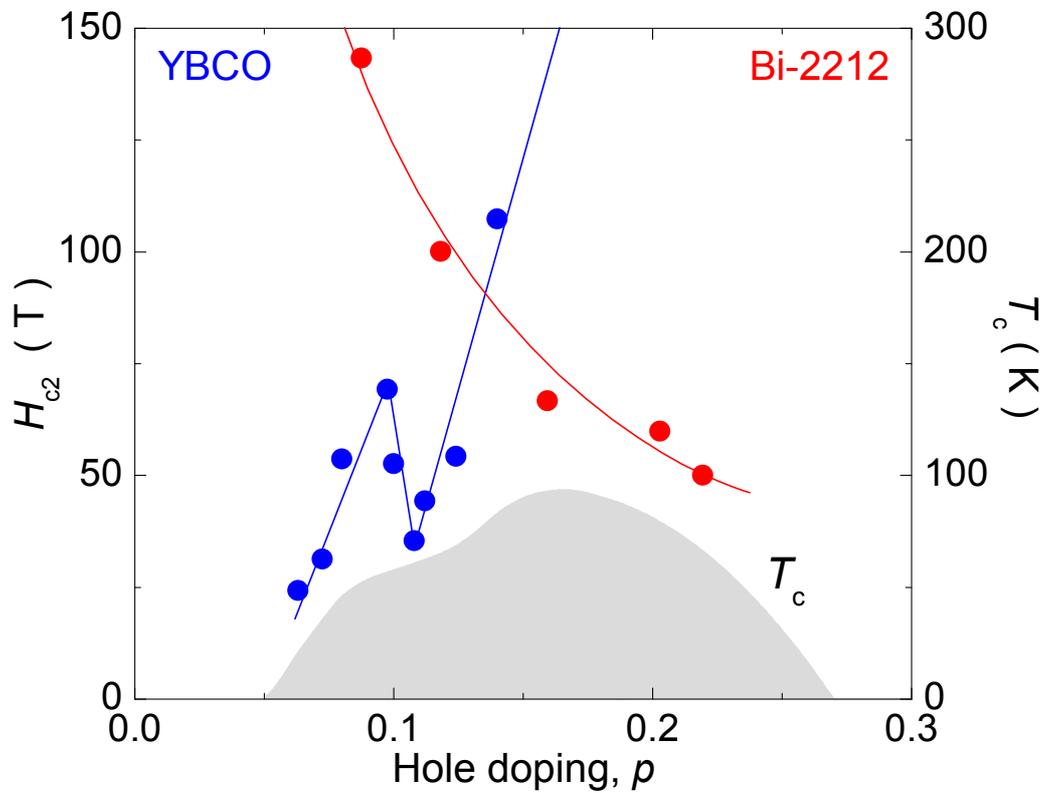

**Figure 1 | Doping dependence of the upper critical field $H_{c2}$.**

Upper critical field $H_{c2}$ of cuprate superconductors vs doping $p$ extracted from magneto-conductivity data on YBCO (ref. 6, blue circles; left axis) and from Nernst data on Bi-2212 (ref. 7, red circles; left axis). These two studies of superconducting fluctuations above $T_c$ led to contradictory conclusions on how the superconducting pairing strength in cuprates varies with doping. The superconducting $T_c$ of YBCO is also shown (ref. 27, grey dome; right axis).

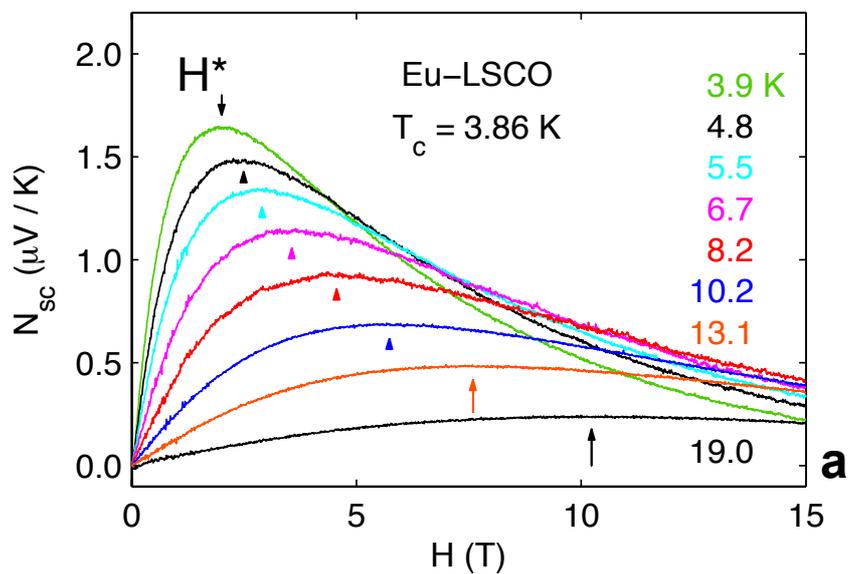

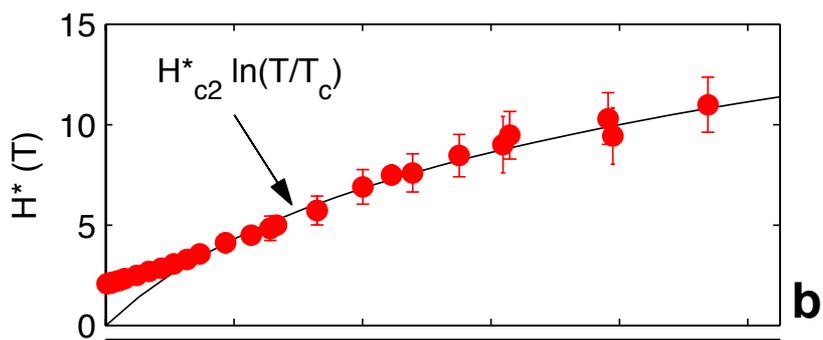

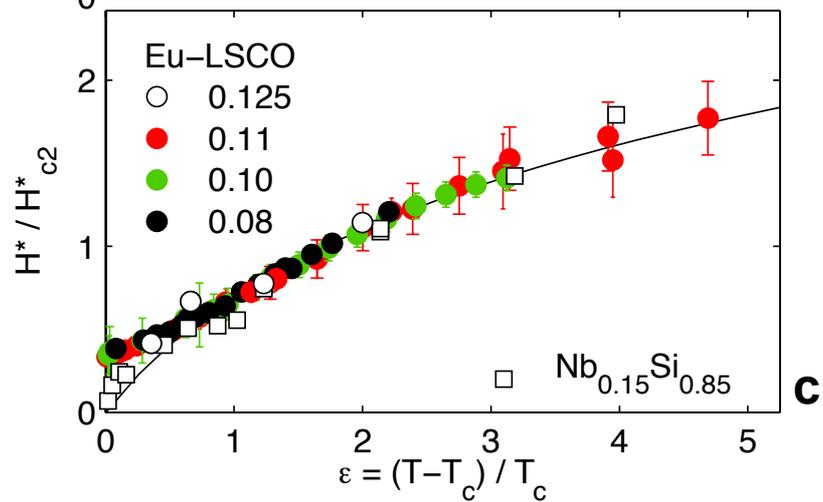

**Figure 2 | Nernst signal in Eu-LSCO and ghost critical field $H^*$.**

**a)** Superconducting Nernst signal $N_{sc}$ of Eu-LSCO plotted as a function of magnetic field $H$, for different values of the temperature $T$ as indicated. The sample is a single crystal with a hole concentration (doping) of $p = 0.11$ and a superconducting transition temperature $T_c = 3.86$ K. The thermal gradient is applied in the $CuO_2$ plane and the transverse Nernst voltage is measured in the perpendicular in-plane direction with the field applied along the c-axis. A quasiparticle background $N_{qp}$ is subtracted from the raw data of $N$ vs $H$ in Fig. S3 (see Fig. S2). The arrows mark the maximum value of $N_{sc} = N - N_{qp}$ for each isotherm, which defines the ghost critical field $H^*$. **b)** Temperature dependence of $H^*$ extracted from the data in a), plotted vs reduced temperature $\varepsilon \equiv (T - T_c) / T_c$. The solid line is a fit to $H^* = H_{c2}^* \ln(T / T_c)$, which provides a model-free field scale, $H_{c2}^*$, equal to 6.2 T for this doping. **c)** Ghost critical field $H^*$ in Eu-LSCO at $p = 0.08, 0.10, 0.11$ and $0.125$, and in the conventional superconductor $Nb_{0.15}Si_{0.85}$ (from ref. 16), plotted as $H^* / H_{c2}^*$ vs $\varepsilon$, with $H_{c2}^*$ obtained as in b) for each sample. The values of $H_{c2}^*$ are given in Table S1. Error bars on $H^*$ in panels b) and c) come from the uncertainty in $N_{qp}$ (Figs. S1 and S2).

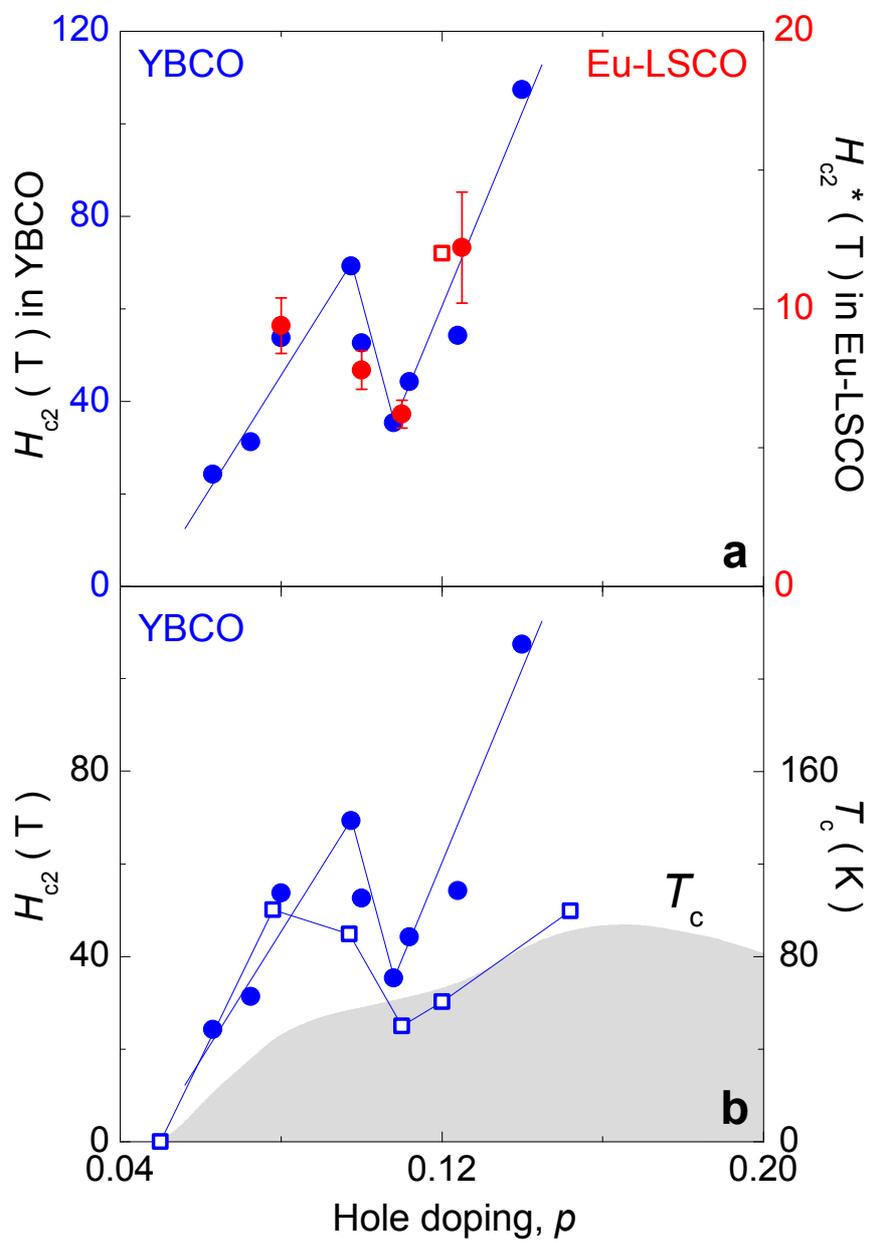

**Figure 3 | Doping dependence of $H_{c2}$ in YBCO and Eu-LSCO.**

**a)** Upper critical field $H_{c2}$ obtained from the superconducting fluctuations above $T_c$, plotted as a function of hole doping $p$. For YBCO (blue symbols; left axis), $H_{c2}$ is defined via the relation $H_{c2} \equiv \Phi_0 / 2\pi\xi_0^2$, in terms of the zero-temperature coherence length $\xi_0$ obtained from a Gaussian analysis of the fluctuation magneto-conductivity (ref. 6). For Eu-LSCO (red symbols; right axis), $H_{c2}$ is taken to be $H_{c2}^*$ in the fit of the ghost critical field $H^*$ to $H^* = H_{c2}^* \ln(T / T_c)$ (Fig. 2). Error bars on $H_{c2}^*$ correspond to the uncertainty in fitting to the $H^*$ data points in Fig. 2c. The red square marks the value of $H_{c2}$ obtained directly from resistivity measurements at $T \to 0$ on Nd-LSCO, a material very similar to Eu-LSCO (Fig. S11). **b)** Comparison of $H_{c2}$ in YBCO determined in two different ways: from $\xi_0$ above $T_c$, as in a), and from high-field transport measurements[10,11] that suppress superconductivity at low temperature ($T \ll T_c$). The two measures of $H_{c2}$ are in reasonable agreement; in particular, they both have a minimum at $p = 0.11$, where $H_{c2} \approx 30$ T. Low non-monotonic values of $H_{c2}$ vs $p$ are also obtained from low-field μSR measurements of the vortex core radius at $T \ll T_c$ (ref. 9). The doping dependence of the zero-field $T_c$ is shown for comparison (grey dome, right axis; ref. 27).

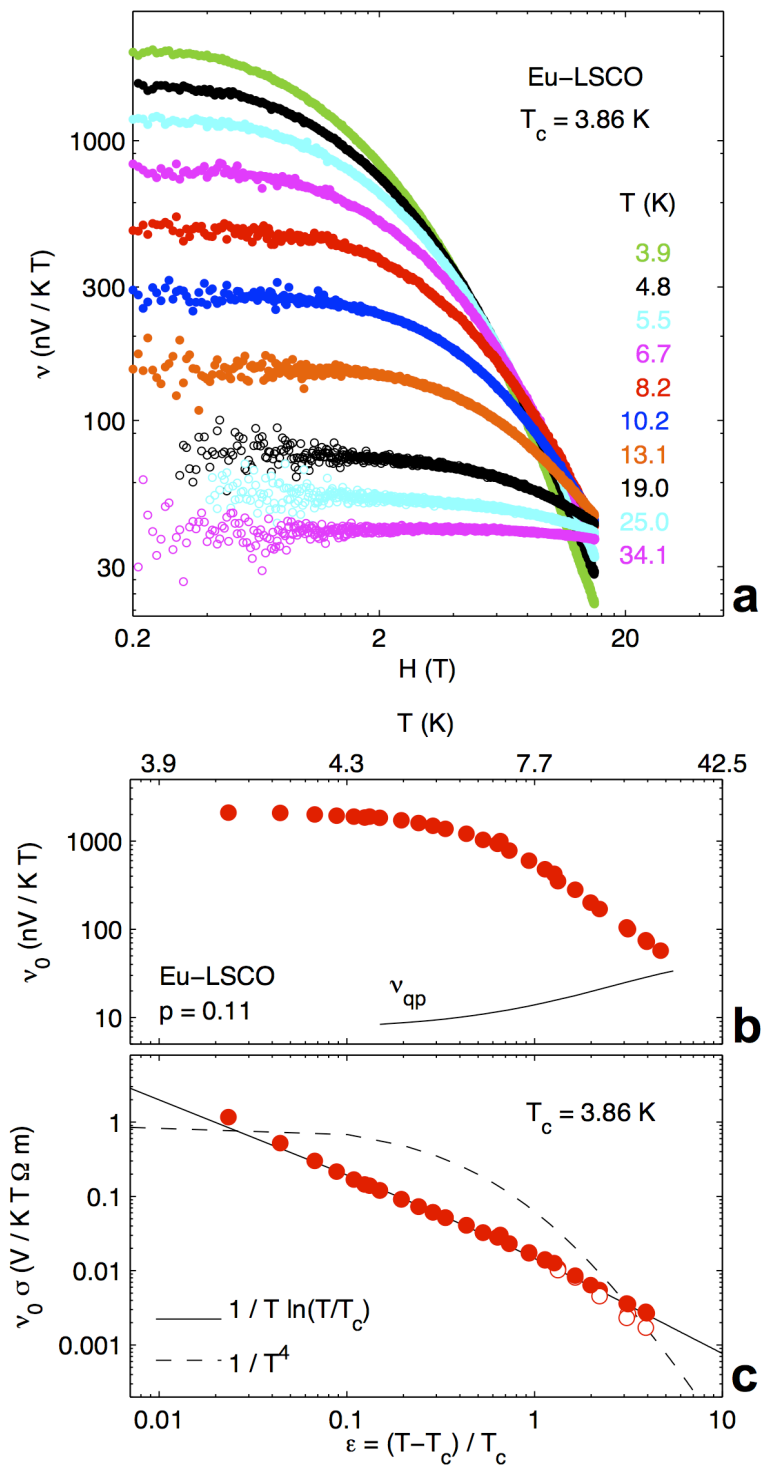

**Figure 4 | Comparison to Gaussian theory.**

**a)** Nernst coefficient $v$ of Eu-LSCO at $p = 0.11$ as a function of magnetic field $H$, plotted on a log-log scale for different values of the temperature $T$ as indicated. **b)** Temperature dependence of the Nernst coefficient in the $H = 0$ limit, $v_0$, plotted vs reduced temperature $\varepsilon$ on a log-log scale. The solid line is the quasiparticle background $v_{qp}$ obtained by suppressing superconductivity with a large magnetic field (Fig. S1). **c)** Same data as in b), multiplied by the zero-field conductivity $\sigma$ of the sample (Fig. S6). Open circles show the effect of subtracting $v_{qp}$ from $v_0$, to obtain $v_{sc}(B\to 0) = v_0 - v_{qp}$. The solid line is the theoretical expectation for the Peltier coefficient in the $B\to 0$ limit, $\alpha_{xy}^{sc} \sim v_{sc}\,\sigma$, from Gaussian fluctuations of the order parameter in a 2D superconductor[19,20]. The dashed line is the theoretical expectation for phase-only fluctuations[21].